\documentclass[twocolumn,nofootinbib,amsmath,amssymb,aps,prd,balancelastpage,superscriptaddress]{revtex4-1}

\usepackage{graphicx}
\usepackage{epsfig,latexsym}

\usepackage[usenames]{color}
\RequirePackage{xspace} \allowdisplaybreaks

\usepackage{natbib}
\usepackage{url}
\usepackage{lineno}

\usepackage{bm}
\usepackage{dcolumn}
\usepackage{graphicx}
\usepackage{graphics}
\usepackage[latin1]{inputenc}
\usepackage{latexsym}
\usepackage{rotating}
\usepackage{dsfont}
\usepackage{xspace} 

\usepackage{ulem}
\usepackage{outlines}
\usepackage{enumitem}
\setenumerate[1]{label=\Roman*.}
\setenumerate[2]{label=\Alph*.}
\setenumerate[3]{label=\roman*.}
\setenumerate[4]{label=\alph*.}

\definecolor {darkgreen}{rgb}{0.2,0.7,0.2}

\usepackage{tikz}
\usetikzlibrary{decorations}
\usetikzlibrary{decorations.text}
\usetikzlibrary{arrows}
\tikzset{l/.style={draw=black, line width=1pt}}
\tikzset{s/.style={dashed, draw=black, line width=0.2pt}}

\begin{document}

\newcommand{\nl}{\nonumber\\}

\newcommand{\ans}{ansatz }
\newcommand{\mat}[4]{\left(\begin{array}{cc}{#1}&{#2}\\{#3}&{#4}
\end{array}\right)}
\newcommand{\matr}[9]{\left(\begin{array}{ccc}{#1}&{#2}&{#3}\\
{#4}&{#5}&{#6}\\{#7}&{#8}&{#9}\end{array}\right)}
\newcommand{\matrr}[6]{\left(\begin{array}{cc}{#1}&{#2}\\
{#3}&{#4}\\{#5}&{#6}\end{array}\right)}
\newcommand{\cvb}[3]{#1^{#2}_{#3}}
\def\lsim{\raise0.3ex\hbox{$\;<$\kern-0.75em\raise-1.1ex
e\hbox{$\sim\;$}}}
\def\gsim{\raise0.3ex\hbox{$\;>$\kern-0.75em\raise-1.1ex
\hbox{$\sim\;$}}}
\def\abs#1{\left| #1\right|}
\def\simlt{\mathrel{\lower2.5pt\vbox{\lineskip=0pt\baselineskip=0pt
           \hbox{$<$}\hbox{$\sim$}}}}
\def\simgt{\mathrel{\lower2.5pt\vbox{\lineskip=0pt\baselineskip=0pt
           \hbox{$>$}\hbox{$\sim$}}}}
\def\unity{{\hbox{1\kern-.8mm l}}}
\newcommand{\eps}{\varepsilon}
\def\ep{\epsilon}
\def\ga{\gamma}
\def\Ga{\Gamma}
\def\om{\omega}
\def\omp{{\omega^\prime}}
\def\Om{\Omega}
\def\la{\lambda}
\def\La{\Lambda}
\def\al{\alpha}
\newcommand{\ov}{\overline}
\renewcommand{\to}{\rightarrow}
\renewcommand{\vec}[1]{\mathbf{#1}}
\newcommand{\vect}[1]{\mbox{\boldmath$#1$}}
\def\tm{{\widetilde{m}}}
\def\mcirc{{\stackrel{o}{m}}}
\newcommand{\Dm}{\Delta m}
\newcommand{\dm}{\varepsilon}
\newcommand{\tanb}{\tan\beta}
\newcommand{\nbar}{\tilde{n}}
\newcommand\PM[1]{\begin{pmatrix}#1\end{pmatrix}}
\newcommand{\up}{\uparrow}
\newcommand{\down}{\downarrow}
\def\omE{\omega_{\rm Ter}}
%

\newcommand{\Dsusy}{{susy \hspace{-9.4pt} \slash}\;}
\newcommand{\DCP}{{CP \hspace{-7.4pt} \slash}\;}
\newcommand{\mc}{\mathcal}
\newcommand{\gr}{\mathbf}
\renewcommand{\to}{\rightarrow}
\newcommand{\gtc}{\mathfrak}
\newcommand{\wh}{\widehat}
\newcommand{\br}{\langle}
\newcommand{\kt}{\rangle}

\newcommand{\Pl}{{\mbox{\tiny Pl}}}
\newcommand{\stat}{{\mbox{\tiny stat}}}
\newcommand{\tot}{{\mbox{\tiny tot}}}
\newcommand{\sys}{{\mbox{\tiny sys}}}
\newcommand{\GW}{{\mbox{\tiny GW}}}
\newcommand{\ny}[1]{\textcolor{blue}{\it{\textbf{ny: #1}}} }
\newcommand{\am}[1]{\textcolor{red}{\it{\textbf{am: #1}}} }

\newcommand{\Hor}{{\mbox{\tiny H}}}
\newcommand{\BH}{{\mbox{\tiny BH}}}
\newcommand{\HL}{{\mbox{\tiny HL}}}
\newcommand{\Bondi}{{\mbox{\tiny Bondi}}}
\newcommand{\DM}{{\mbox{\tiny DM}}}
\newcommand{\Rel}{{\mbox{\tiny Rel}}}
\newcommand{\IGM}{{\mbox{\tiny IGM}}}
\newcommand{\ISM}{{\mbox{\tiny ISM}}}
\newcommand{\CMB}{{\mbox{\tiny CMB}}}
\newcommand{\DE}{{\mbox{\tiny DE}}}
\newcommand{\tidal}{{\mbox{\tiny tidal}}}
\newcommand{\nonspin}{{\mbox{\tiny no-spin}}}
\newcommand{\spinalign}{{\mbox{\tiny spin-aligned}}}
\newcommand{\comp}{{\mbox{\tiny comp}}}


\def\lsim{\mathrel{\mathop  {\hbox{\lower0.5ex\hbox{$\sim$}
\kern-0.8em\lower-0.7ex\hbox{$<$}}}}}
\def\gsim{\mathrel{\mathop  {\hbox{\lower0.5ex\hbox{$\sim$}
\kern-0.8em\lower-0.7ex\hbox{$>$}}}}}

\def\nn{\\  \nonumber}
\def\de{\partial}
\def\brf{{\mathbf f}}
\def\bbf{\bar{\bf f}}
\def\bF{{\bf F}}
\def\bbF{\bar{\bf F}}
\def\bA{{\mathbf A}}
\def\bB{{\mathbf B}}
\def\bG{{\mathbf G}}
\def\bI{{\mathbf I}}
\def\bM{{\mathbf M}}
\def\bY{{\mathbf Y}}
\def\bX{{\mathbf X}}
\def\bS{{\mathbf S}}
\def\bb{{\mathbf b}}
\def\bh{{\mathbf h}}
\def\bg{{\mathbf g}}
\def\bla{{\mathbf \la}}
\def\bmu{\mathbf m }
\def\by{{\mathbf y}}
\def\bmu{\mbox{\boldmath $\mu$} }
\def\bsig{\mbox{\boldmath $\sigma$} }
\def\bunity{{\mathbf 1}}
\def\cA{{\cal A}}
\def\cB{{\cal B}}
\def\cC{{\cal C}}
\def\cD{{\cal D}}
\def\cF{{\cal F}}
\def\cG{{\cal G}}
\def\cH{{\cal H}}
\def\cI{{\cal I}}
\def\cL{{\cal L}}
\def\cN{{\cal N}}
\def\cM{{\cal M}}
\def\cO{{\cal O}}
\def\cR{{\cal R}}
\def\cS{{\cal S}}
\def\cT{{\cal T}}
\def\eV{{\rm eV}}
%

\title{New Massive JT Multi-Gravity \& N-Replica of SYK Models}

\author{Andrea Addazi}
\email{addazi@scu.edu.cn}
\affiliation{College of Physics Sichuan University Chengdu, 610065, China}
\affiliation{INFN sezione Roma {\it Tor Vergata}, I-00133 Rome, Italy, EU}
\author{Jakub Bilski}
\email{bilski@zjut.edu.cn}
\affiliation{Institute for Theoretical Physics and Cosmology, Zhejiang University of Technology, 310023 Hangzhou, China}
\author{Qingyu Gan}
\email{gqy@stu.scu.edu.cn}
\affiliation{College of Physics Sichuan University Chengdu, 610065, China}
\author{Antonino Marcian\`o}
\email{marciano@fudan.edu.cn}
\affiliation{Center for Field Theory and Particle Physics \& Department of Physics, Fudan University, 200433 Shanghai, China}
\affiliation{Laboratori Nazionali di Frascati INFN, Frascati (Rome), Italy, EU}

\begin{abstract}
\noindent
We study a series of powerful correspondences among new multi-gravity extensions of the Jackiw-Teitelboim model, multi-SYK models and multi-Schwarzian quantum mechanics, in the $\rm{(A)dS_{2}/CFT}$ arena. Deploying a $BF$-like formulation of the model, we discuss the counting of the degrees of freedom for some specific classes of multi-gravity potentials, and unveil connections among a variety of apparently different models. Quantization of multi-gravity models can be then achieved from both the Hartle-Hawking no-boundary proposal, the SYK partition function and the spin-foam approaches. We comment on the SYK quantization procedure, and deepen in the appendix the quantization scheme naturally achieved in the $BF$ framework. The new multi-gravity theory hence recovered presents intriguing applications for analogue gravitational models developed for condensed matter physics, including graphene, endowed with defects and high intensity magnetic fields.

\end{abstract}

\maketitle

\vspace{0.3cm}
\section{Introduction} 
\noindent 
Recent studies \cite{N1,N2,N3,N4,N5,Saad:2019lba,Maldacena:2019cbz,Iliesiu:2019xuh} on the Jackiw-Teitelboim (JT) $1\!+\!1D$ gravity \cite{J,T,P} in $(A)dS_{2}$ have shown its surprising duality with $1D$ Schwarzian quantum mechanics and the Sachdev-Ye-Kitaev model \cite{SY,K1,K2} (SYK). At the same time, the urgent question arises whether a consistent theory of massive gravity could be naturally formulated. This naturally leads to a concept of multi-gravity models, where multi-parallel metrics are coupled by an interaction potential --- see e.g. Refs.~\cite{deRham:2010ik,deRham:2010kj,Hassan:2011hr}.\\

In this letter, we show the existence of a new class of $1\!+\!1D$ $BF$-like theories coupled to the dilaton fields, each one equivalent to $1\!+\!1D$ gravity, which can be connected to $N$-copies of mutually coupled SYK fermionic models. In turn, it is well known that $\rm{dS_{2}}$ theories are holographically connected to $1D$ conformal field theories $CFT_{1}\times CFT_{1}'$. Therefore, in our case, a large class of multi-gravity $1\!+\!1D$ models coupled to $N$ dilatons is described by $N$ copies of the $CFT_{1}$ theory. Our current analysis is strongly motivated by the purpose of modeling graphene layers with lattice defects (from a sample of carbon atoms removed), in presence of high magnetic fields or in superconductive phases \cite{G0,G1,G2}. The dynamics of fermions on a planar graphene with topological defects and a local curvature was deepened is a series of papers, including the seminal study in Refs.~\cite{CM}. The dynamics of the graphene layers was not taken into account so far. Nonetheless, the simplified dual $1\!+\!1D$ multi-gravity framework we are considering may allow to capture new non-perturbative aspects, which are testable in graphene physics.

\section{The model} 
\noindent
We start our analysis moving from the JT 1+1\emph{D} gravity action in the presence of a boundary term, namely
\begin{align}
\begin{split}
\mathcal{S}=&\;\phi^0\bigg[\int_{\Sigma}  d^{2}x \sqrt{-g}(R-2)-2\int_{\partial \Sigma}  K \bigg]
\\
\label{iS}
&+\int_{\Sigma} d^{2}x  \sqrt{-g} \phi (R-2)-2\int_{\partial \Sigma}  \phi^b\, K\,,
\end{split}
\end{align}
where the cosmological constant is normalized as $\Lambda=1$. Here, $\Sigma$ is the $1\!+\!1D$ bulk, $\partial \Sigma$ is its boundary, $\phi$ is the dilaton field, $\phi^0$ is a constant, $\phi^b$ is the dilaton on the boundary, $R$ denotes the Ricci scalar in the bulk and $K$ is the extrinsic scalar curvature on the boundary. The boundary $\partial \Sigma$ has the topology of a circle, while the bulk coincides with $dS_{2}$ metric. It is worth to note that the first term in the action is completely topological. The theory becomes dynamical only because of the presence of the dilaton, which is here the only propagating degree of freedom.
Adopting the same jargon as in Ref.~\cite{N2}, we can call this latter a nearly $dS_2$ background. We can then implement a $BF$-like formulation, with a dilaton coupled to the $BF$ system. The action for the bulk then becomes
\begin{equation}
\label{iS3}
\mathcal{S}_{\Sigma}=\phi^0\!\int_\Sigma {\rm Tr} \left[ B \wedge F[A]\right]+\int_\Sigma  \phi \, {\rm Tr} \left[  B \wedge F[A] \right]\,,
\end{equation}
where in $1\!+\!1D$ the term Tr[$B \wedge F$] trivialises to Tr[$BF$] --- the trace ${\rm Tr[...]}$ is meant to be normalized to unity. For simplicity of notation, we omit to write the trace in what follows. The field-strength of the index-less SO$(1,1)$ connection\footnote{In $dS_{2}$ we can promote the SO$(1,1)$ group to SO$(2,1)$, or equivalently to SL$(2,\mathbb{R})$, as explained in Appendix A (see also Ref.~\cite{Cotler:2019nbi,Grumiller:2020elf}).  } $A$ is denoted by $F[A]= \partial_\mu A_\nu \, dx^{\mu} \wedge dx^{\nu}$, and the $B$-form reduces to an index-less 0-form. Imposing on-shell gauge-invariance, the boundary term casts
\begin{equation}
\label{iS2}
\mathcal{S}_{\partial \Sigma}=\phi^0\!\int_{\partial \Sigma} B^b \wedge A^b+\int_{\partial \Sigma}\phi^b\, B^b \wedge A^b\,,
\end{equation}
where again $B^b \wedge A^b$ trivializes to $(B^b \, A^b)$, and $A^b$ denotes the pull-back of the connection on $\partial \Sigma$. In the bulk, the dynamics of the system is easily calculable. Varying with respect to $\phi$, $B$ and $A$, one finds
\begin{equation}
\label{E1}
B(\partial_{\mu}A_{\nu}\epsilon^{\mu\nu})=0 \  \ \longrightarrow \ \ B\neq 0, \ \ dA=0\, ,
\end{equation}
\begin{eqnarray}
\label{E2}
\!\!\! \phi= -\phi^0, \ \  dA\neq 0 \ \
{\rm or}\ \  dA=0,\ \  \phi \,\,\, {\rm unconstrained},
\end{eqnarray}
\begin{equation}
\label{E3}
\phi^0\,dB+d(\phi B)=0 \ \ \longrightarrow \ \  B=\frac{C_1}{\phi+\phi^0}\,,
\end{equation}
$C_1$ being an integration constant. Without loss of generality, in what follows we can simplify solution \eqref{E3}, setting $C_1=1$ . This system of equations implies that $B=1/\phi$, having set $\phi^0=0$, and thus the only propagating degree of freedom (DOF) is described by the theory in the bulk. The boundary terms do not alter the DOFs counting. 

\section{Interacting DOFs}  
\noindent
We can now consider a multi-$BF$-like extension of the previous example. Consequently we introduce $N$ parallel sets of $\{B_{i},A_{i},\phi_{i}\}$ fields, with $i=1,...N$, generically coupled through a potential $\mathcal{V}[\phi_{1},..,\phi_{n},B_{1},..,B_{n}]$. The number of propagating DOFs depends on the specific form of the potential. For instance, adding a potential $\mathcal{V}$ to the action on the bulk in Eq.~\eqref{iS3}, the new equations of motion become
\begin{equation}
\label{Eo1}
B^i(\partial_{\mu}A_{i \nu}\epsilon^{\mu\nu})= -\frac{\delta \mathcal{V}}{\delta\phi_i} 
\, ,
\end{equation}
\begin{eqnarray}
\label{Eo2}
(\partial_{\mu}A_{i \nu}\epsilon^{\mu\nu}) (\phi_i+ \phi_i^0) = - \frac{\delta \mathcal{V}}{\delta B_i}\,,
\end{eqnarray}
\begin{equation}
\label{Eo3}
\phi^0_i\,dB_i+d(\phi_i B_i)= 0 \,.
\end{equation}
Considering a generic potential $\mathcal{V}[B_i, B_j,\phi_i, \phi_j]$, two possible cases are recovered: i) the theory continues to imposes the flatness of the connections $A^i$, whenever the potential satisfies
\begin{equation}
\label{flat}
 \frac{\delta \mathcal{V}}{\delta B_i}=\frac{\delta \mathcal{V}}{\delta\phi_i}=0\,,
\end{equation}
thus reducing the potential to trivial cases; ii) whenever the potential is not assumed to be trivial, extremizing conditions can be imposed that allows to find configurations that entail flat connections. \\

Interesting examples of potential are provided by a simple polynomial choice, \textit{i.e.}
\begin{equation}
\label{extremeflat}
\mathcal{V}[B_i, B_j,\phi_i, \phi_j]=\left( \sum \limits_{i=1}^{N} \alpha_i B_i +\frac{\beta_i} {\phi_i+\phi^0_i} \right)^N\,.
\end{equation}
This choice can be consistent with Eq.~\eqref{Eo3}. Furthermore, it naturally provides massive terms for the \emph{1+1D} gravitational sectors, and self-interactions among each sector of $B$ frame-fields and $\phi$ dilaton fields, and among the two sectors.  

\section{M\"obius mapping of DOFs}
\noindent
There exist also potentials that are not compatible with having propagating DOFs arising from the minimal kinetic term in Eq.~\eqref{iS3}. This class of potentials indeed realizes a one-to-one identification of the $B$-fields with dilatons the $\phi$-fields that is different than $B_{i}=1/(\phi_{i}+\phi_i^0)$. Therefore imposing Eq.~\eqref{Eo3} results in freezing completely the DOFs.

An illustrative example is provided by the potential of a non-trivial bi-gravity/bi-dilaton case,
\begin{equation}
\begin{split}
\label{BB}
\mathcal{V}= & \left[c\alpha B_{1}\left(\alpha \phi_{1}+\beta \phi_{2}\right)+d\alpha B_{1}-a\alpha \phi_{1}-\frac {b}{2}\right] \\
& \times \left[c\beta B_{2}\left(\alpha \phi_{1}+\beta \phi_{2}\right)+d\beta B_{2}-a\beta \phi_{2}-\frac {b}{2}\right]\,,
\end{split}
\end{equation}
with constants $\alpha,\beta,a,b,c,d$ satisfying $ad-bc=1$.
A specific branch of solutions corresponds to two DOFs that mix according to
\begin{equation}
\label{EoMA}
 c\alpha B_{1}\left(\alpha \phi_{1}+\beta \phi_{2}\right)+d\alpha B_{1}-a\alpha \phi_{1}-\frac {b}{2}=0\, ,
\end{equation}
\begin{equation}
\label{EoMB}
c\beta B_{2}\left(\alpha \phi_{1}+\beta \phi_{2}\right)+d\beta B_{2}-a\beta \phi_{2}-\frac {b}{2}=0\, ,
\end{equation}
$A_{1,2}$ remaining flat on these directions that extremize the potential. Solving the system in $\{B_{1},B_{2}\}$, in terms of $\{\phi_{1}, \phi_{2}\}$, we obtain
\begin{equation} \nonumber
B_{1}=\frac{a \alpha \phi_{1}+b / 2}{c \alpha\left(\alpha \phi_{1}+\beta \phi_{2}\right)+d \alpha},
\end{equation}
\begin{equation} \label{Sol}
B_{2}=\frac{a \beta \phi_{2}+b / 2}{c \beta\left(\alpha \phi_{1}+\beta \phi_{2}\right)+d \beta}\, .
\end{equation}
This indeed implies that the number of propagating DOFs are reduced to two ones, recovered out of the initial four unconstrained fields. It is interesting to note that  linear combinations $\alpha B_{1}+\beta B_{2}$ and $\alpha \phi_{1}+\beta \phi_{2}$ relate to one another according to the SL$(2,\mathbb{R})$ map, namely
\begin{equation}
\label{B12}
\alpha B_{1}+\beta B_{2}=\frac{a\left(\alpha \phi_{1}+\beta \phi_{2}\right)+b}{c\left(\alpha \phi_{1}+\beta \phi_{2}\right)+d}, \quad a d-b c=1.
\end{equation}
 \\

There exists a systemic way to construct a large class of potentials to realize the M\" obius map between $B$ and $\phi$ fields. We consider any potential that can be decomposed as
\begin{equation}
\mathcal{V}=\mathcal{V}^m_1 \mathcal{V}^n_2\,,
\end{equation}
and satisfies
\begin{equation}
\begin{split}
\mathcal{V}_1 + \mathcal{V}_2 = & c\left(\alpha B_{1}+\beta B_{2}\right)\left(\alpha \phi_{1}+\beta \phi_{2}\right)+d\left(\alpha B_{1}+\beta B_{2}\right) \\
&  -a\left(\alpha \phi_{1}+\beta \phi_{2}\right)-b, \quad  ad-bc=1,
\end{split}
\end{equation}
where $m,n$ are positive integers. Applying the extreme condition Eq.~(\ref{flat})  yields $\mathcal{V}_1=0$ and  $\mathcal{V}_2=0$, which reshuffles into $\mathcal{V}_1+\mathcal{V}_2=0$, namely Eq. (\ref{B12}). Therefore, the potential constructed in this way leads to the SL$(2,\mathbb{R})$ map among $B_{1,2}$ and $\phi_{1,2}$ fields for the flat connection $dA_{1,2}=0$. In particular, if one takes $\mathcal{V}_{2}=0$, then  a trivial potential can be constructed as $\mathcal{V}=\mathcal{V}^m_1$ with integer $m>1$.\\

It is straightforward to generalize the bi-gravity/bi-dilaton case to the multi-gravity/multi-dilaton one, reiterating the constructive approach hitherto specified. Similarly, the potentials that are  required for this purpose can be rewritten as 
\begin{equation}
\mathcal{V}=\mathcal{V}_1^{m_1} \mathcal{V}_2^{m_2} \cdots \mathcal{V}_n^{m_n}
\end{equation}
with $m_1,m_2,\cdots, m_n \in \mathbb {N}^+$ and such that 
\begin{equation}
\mathcal{V}_1+ \mathcal{V}_2+ \cdots +\mathcal{V}_n =  c \sum \nolimits _{i}  \alpha_{i} B_{i} \sum \nolimits _{j} \alpha_{j} \phi_{j} \ +
\end{equation} 
\begin{equation}
\ \ \qquad \qquad \qquad \qquad \qquad  d \sum \nolimits _{i} \alpha_{i} B_{i}   -a  \sum \nolimits _{j} \alpha_{j} \phi_{j} -b  \nonumber \,,
\end{equation} 
with the constraint $ad-bc=1$, and $\alpha_i$ constants (for $i=1,2,\cdots, N$) --- notice that $N$ does not necessarily equal $n$.  Consequently, $A_{i}$ remains flat on the direction where all multiplicative factors $\mathcal{V}_i (i=1,2,\cdots, n)$ vanish, implying that the mapping among the linear combination $\sum \nolimits _{i} \alpha_{i} B_{i}$ and $\sum \nolimits _{i} \alpha_{i} \phi_{i}$ instantiates a M\"obius transformation. Different potentials that fall within this same class, endowed with different parameters, span all the possible transformations of the SL$(2,\mathbb{R})$ group. Finally, imposing Eq.~\eqref{Eo3} freezes all the DOFs of the system.

\section{Deformed kinematics with M\"obius map} 
\noindent
We may prevent to freeze all the DOFs of the system, moving from a slightly different kinetic term in the action. For instance, we may consider a simple non-trivial multi-gravity/multi-dilaton model specified by the action on the bulk
\begin{equation}
\label{SN2}
\mathcal{S}_{\Sigma}=\sum \limits_{i=1}^{N} f_i^0\!\int\!B_i \wedge F(A_i)+\!\int\!f_i \, B_i \wedge F(A_i)\,,
\end{equation}
provided that a generic functional $f[\phi_i; \phi_j]:=f_i$ is chosen that shows a dependence on the $\phi_i$ multiplet of dilaton-fields, and constant $f_i^0$. To be compatible with the discussion in last section, we consider a simple realization as
\begin{align}
\label{effei}
f_i=\frac{c \alpha_i \left(\sum \nolimits_{j=1}^{N}\alpha_{j} \phi_{j}\right)+d \alpha_i}{a \alpha_i \phi_{i}+b / N}.
\end{align}

It is straightforward to realize that the variation with respect to the connections $A_i$ of the action in Eq.~\eqref{SN2} provides
\begin{eqnarray}
\label{E3N}
&&f_i^0\,dB_i+d\big(f_i B_i\big)=0 \nonumber \\
&& \qquad \! \quad \quad \longrightarrow \ \  B_i=\frac{C_{i1}}{(f_i +f^0_i)}\,.
\end{eqnarray}

We can then fix the constants as previously done, setting  $C_{i1}\!=\!1$. This new equation of motion is now compatible with the extremal conditions for the potentials $\mathcal{V}$, proving dynamical degrees of freedom.\\

In the simplest bi-gravity/bi-dilaton model, the action on the bulk provides, for a choice of the $f_i$ functional as specified in Eq.~\eqref{effei}, a relation between the DOFs of the gravity sector that is implemented through the M\"obius map, \textit{i.e.}

\begin{align}
\begin{split}
\label{SN2bi}
\mathcal{S}_{\Sigma}=& \!\int\! \left( \phi_1^0+ \frac{c \, \alpha\left(\alpha \phi_{1}+\beta \phi_{2}\right)+d \alpha}{a \alpha \phi_{1}+b / 2} \right)  B_1 \wedge F[A_1]   \\
&+\!\int\! \left( \phi_2^0+ \frac{c \beta \left(\alpha \phi_{1}+\beta \phi_{2}\right)+d \beta}{a \beta \phi_{2}+b / 2} \right)  B_2 \wedge F[A_2] .
\end{split}
\end{align}
Varying with respect to the fields $\phi_{1,2}$, $B_{1,2}$, this latter action respectively provides:
\begin{equation}
\begin{aligned}
\left( c (\alpha \phi_{1}+\beta  \phi_{2}) +d\right) B_{1} \partial_{\mu} A_{1 \nu} \epsilon^{\mu \nu} &=0, \\
\left(c (\alpha \phi_{1}+\beta  \phi_{2})+d\right) B_{2} \partial_{\mu} A_{2 \nu} \epsilon^{\mu \nu} &=0, \\
 \left( c (\alpha \phi_{1}+\beta  \phi_{2}) +d\right) \partial_{\mu} A_{1 \nu} \epsilon^{\mu \nu} &=0, \\
\left( c (\alpha \phi_{1}+\beta  \phi_{2}) +d\right) \partial_{\mu} A_{2 \nu} \epsilon^{\mu \nu} &=0,
\end{aligned}
\end{equation}
where $\phi^0_{1,2}$ are set to zero. Combining with the equations of motion of $B$ provided in Eq. (\ref{E3N}), namely
\begin{equation}
\begin{aligned}
B_{1} &= \frac{a \alpha \phi_{1}+b / 2}{\alpha  (c \left(\alpha \phi_{1}+\beta \phi_{2}\right)+d \alpha}, \\
B_{2} &= \frac{a \beta \phi_{2}+b / 2}{c \beta\left(\alpha \phi_{1}+\beta \phi_{2}\right)+d \beta},\\
\end{aligned}
\end{equation}
one can see that to have well-defined $B$ fields, we must impose $c (\alpha \phi_{1}+\beta  \phi_{2})+d \neq 0$, which implies $dA_{1,2}=0$ under the imposition of the extremal conditions.
Therefore, a generic SL$(2,\mathds{R})$ transformation is achieved by considering the linear combination $\alpha B_1+\beta B_2$. Moreover, these equations of motion unfreeze two propagating DOFs for the dynamical dilaton $\phi_{1,2}$-fields. 

\section{Boundary terms and Schwarzian theory} 
\noindent 
The boundary terms of $N$-gravity, corresponding to the field configurations $\phi_i$, $B_i[\phi]$ and $A_i$, are expressed on the nearly $dS_{2}$ manifold, for the action in Eq.~\eqref{SN2}, by
\begin{equation}
\label{iS2}
\mathcal{S}_{\partial \Sigma_i}=f^0_i\int_{\partial \Sigma_i}\!\!B^b_i 
A^b_i+\int_{\partial \Sigma_i}\!\!f_i^b \,B^b_i
A^b_i\,,
\end{equation}
where we remind that $f_i:=f[\phi_i; \phi_j]$. \\

Moving back to the JT metric theories, if we consider in Eq.~\eqref{iS2} $f_{i,j}$ rather than $\phi_{i,j}$ as fundamental fields, we obtain the equations of motion 
\begin{equation}
\label{eqq}
\nabla_{\mu}\nabla_{\nu}f_{i}-g_{\mu\nu}\Box f_{i}+g_{\mu\nu}f_{i}=0\,,
\end{equation}
yielding as a solution for the $f_{i}$ the expression
\begin{equation}
\label{ff}
f_{i}=\frac{\alpha_{i}+\gamma_{i}t_{i}+\delta_{i}(t_{i}^{2}+z^{2}_{i})}{z_{i}}\, , 
\end{equation}
corresponding to the dS-coordinates of the $i$-th metrics. It is manifest that the $f_{i}$ are diverging close to the boundaries. We can then relate the $f_{i}$ to the renormalized non-singular fields, namely 
\begin{equation}
\label{fff}
f^{b}_{i}=\frac{f_{i\, r}}{\epsilon}\, , 
\end{equation}
where $f_{i\, R}$ remains constant in the limit $\epsilon\rightarrow 0$. \\

This is an important difference among the single-sector JT theory and the multi-gravity model we are focusing on: we need to impose a renormalization condition on $f_{i}^{b}$ rather than on the multi-dilatons $\phi_{i}^{b}$. Indirectly, this condition amounts to a complicated system of multi-dilaton equations cast in term of the renormalized $f_{i,\, r}^{b}$ fields and $\epsilon$. 

In particular, we can determine the renormalized fields, and figure out that these correspond to 
\begin{equation}
\label{ska}
f_{i\,r}=\frac{\alpha_{i}+\gamma_{i}t_{i}(u_{i})+\delta_{i}t_{i}(u_{i})^{2}}{t'_{i}(u_{i})}\, . 
\end{equation}
The multi-JT-action, and equivalently the corresponding multi-BF, can be rewritten 
in terms of the non-singular fields as follows 
\begin{equation}
\label{rewritten}
-\int\frac{du_{i}}{\epsilon} \frac{f_{i\,r}(u_{i})}{\epsilon}K\,,
\end{equation}
and $K$ expanded in terms of $\epsilon$ as 
\begin{equation}
\label{Kk}
K=\frac{t'_{i}(t_{i}'^{2}+z'^{2}_{i}+z_{i}z''_{i})-z_{i}z'_{i}t''_{i}}{(t'^{2}+z'^{2})^{3/2}}=1+\epsilon^{2}{\rm Sch}(t_{i},u_{i})\, , 
\end{equation}
where $K$ also corresponds to the boundary $[BF]_{b}$, and 
\begin{equation}
\label{II}
{\rm Sch}(t_{i},u_{i})=-\frac{1}{2}\Big(\frac{t''}{t'}\Big)^{2}+\Big( \frac{t''}{t'}\Big)'\, .
\end{equation}
Thus the final action cast in term of the decoupled forms 
\begin{equation}
\label{dec}
I=-\sum_{i=1}^{N}\frac{1}{8\pi G}\int du_{i} f_{i\,r}(u_{i}){\rm Sch}(t_{i},u_{i})\, . 
\end{equation}
The $f_{i\,r}(u_{i})$ are interpreted as external couplings, while $t_{i}(u_{i})$ denotes 
the wave-function. \\

We may focus on the simplest case of bi-dilaton-gravity. Physically, the two dilaton wave-functions are localized on both the boundary circles of the two $dS_{2}$ bulk metrics. On the other hand, their wave-functions can be re-expressed in terms of their non-linear combination as $f_{1,2}$. The $f_{1,2}$ behave as wave-functions separately localized on the first and the second boundary circles respectively, i.e. $f_{1}$ lives on the boundary of the first metric but not on the boundary of the second metric.

Thus the two dilatons, non-linearly and conformally coupled to the two metrics, can be described by two Schwarzian models where the external coupling fields are $f_{1,2}^{r}(u_{1,2})$. It is also worth to note that $f_{1,2}$ are related each others as follows: 
\begin{equation}
\label{ffd}
\frac{f_{1}^{b}}{f_{2}^{b}}=\frac{f_{1}^{r}}{f_{2}^{r}}=\Big(\frac{c\alpha(\alpha \phi_{1}^{b}+\beta\phi_{2}^{b})+d\alpha}{c\beta(\alpha \phi_{1}^{b}+\beta \phi_{2}^{b})+d\beta}\Big)\Big(\frac{a\beta \phi_{2}^{b}+b/2}{a\alpha \phi_{1}^{b}+b/2}\Big)\, ,
\end{equation}
where $\phi_{1,2}^{b}$ correspond to $\phi_{1,2}$, localized on both the boundaries of $dS_{2}$, according to the $f_{1,2}(u_{1,2})$ non linear boundary conditions. 

\section{Relation to the SYK model} 
\noindent
The connection with a \textit{N}-replica of SYK models can be easily recovered while moving from the expression for $N$ uncoupled Schwarzian theories in Eq.~\eqref{dec}. As an illustrative example, we first consider the $N=2$ case. Accounting for the four fermion interaction, we may cast the Hamiltonian as 
\begin{align}
\begin{split}
\label{HH}
\mathcal{H}=&\!\!\sum_{jklm}\!\!J_{jklm}\chi_{j}\chi_{k}\chi_{l}\chi_{m}+
\\
&\!\!\sum_{j'k'l'm'}\!\!\!\!K_{j'k'l'm'}\zeta_{j'}\zeta_{k'}\zeta_{l'}\zeta_{m'}+\mathcal{V}[\chi_{p},\zeta_{q}]\, ,
\end{split}
\end{align}
where the variables $\chi \!=\! \chi(\tau)$ and $\zeta \!=\! \zeta(\tau)$, satisfying $\{\chi_{j},\chi_{k}\}=\delta_{jk}$, $\{\zeta_{j},\zeta_{k}\}=\delta_{jk}$ and $\{\chi_{l},\zeta_{m}\}=0$, have been introduced, and $\mathcal{V}[\chi_{p},\zeta_{q}]$ has been chosen as a generic coupling potential between the two fermions species, up to the fourth order. Furthermore, $\mathcal{J}$ and $\mathcal{K}$ are two random matrices such that their square averaged values acquire the forms $\overline{J_{jklm}^{2}}=3!J^2/N^{3}$ and $\overline{K_{jklm}^{2}}=3!K^2/M^{3}$, with $J$ and $K$ being characteristic energy scales, and $N$ and $M$ numbers of the $\chi$ and $\zeta$ fermionic species. For a large class of potentials, the doubly coupled SYK models can be diagonalized in two states, which are described by two decoupled SYK models. A further simplified example is obtained considering
\begin{align}
\begin{split}
\label{SSYY}
\mathcal{H}=&\;2\!\!\sum_{jklm}\!\!J_{jklm}\chi_{j}\chi_{k}\chi_{l}\chi_{m} \\
&+2\!\!\sum_{j'k'l'm'}\!\!J_{j'k'l'm'}  \zeta_{j'}\zeta_{k'}\zeta_{l'}\zeta_{m'} \\
&+12\!\!\sum_{j''k''l''m''}\!\!J_{j''k''l''m''}\chi_{j''}\chi_{k''}\zeta_{l''}\zeta_{m''}\, ,
\end{split}
\end{align}
where the same number of $\zeta,\chi$ species is assumed. The Hamiltonian in Eq.~\eqref{SSYY} can be easily diagonalized into
\begin{equation}
\label{JJJ}
\mathcal{H}=\sum_{jklm}J_{jklm}\left( \chi_{j}^{+}\chi_{k}^{+}\chi_{l}^{+}\chi_{m}^{+}
+\chi_{j}^{-}\chi_{k}^{-}\chi_{l}^{-} \chi_{m}^{-} \right)\,,
\end{equation}
where $\chi^{\pm}=\chi\pm \zeta$. \\

In the limit of large coupling limit $\beta J\!>\!\!>\!1$ ($\beta=1/T$ denotes the inverse temperature and Euclidean time), the diagonalized model is endowed with propagators of $\chi\pm \zeta$ that are invariant under arbitrary changes of the time coordinate, \textit{i.e.}
\begin{equation}
\label{Gtautau}
G_{\pm}(\tau_{1},\tau_{2})\rightarrow G_{\pm}(\tilde{\tau}_{\pm}(\tau_{1}),\tilde{\tau}_{\pm}(\tau_{2})) \ \tilde{\tau}_{\pm}(\tau_{1})^{\frac{1}{4}}\tilde{\tau}_{\pm}(\tau_{2})^{\frac{1}{4}}\, ,
\end{equation}
where $\tilde{\tau}(\tau)=\tau_{*i}e^{2i\pi \tau/\beta}$, $\tau_{*}$ standing for an arbitrary constant. Time reparametrizations act as conformal maps, and a double decoupled Schwarzian theory emerges. Consequently, a multi-SYK can be easily envisaged, by generalizing the bi-SYK model we discussed above.

In the case of a double Schwarzian models, as in Eq.\eqref{dec} adapted to $N=2$, the two Schwarzian transformations can be related to the definition of the propagators
of two fermions:
\begin{equation}
\label{kjaj}
\langle \psi_{1}(\tau_{1}) \psi_{1}(\tau_{2}) \rangle= G_{1}(\tau_{1},\tau_{2})\rightarrow G_{1}(\tilde{\tau}_{1}(\tau_{1}),\tilde{\tau}_{1}(\tau_{2}))\, ,
\end{equation}
\begin{equation}
\label{kjaj}
\langle \psi_{2}(\tau_{1}) \psi_{2}(\tau_{2}) \rangle= G_{2}(\tau_{1},\tau_{2})\rightarrow G_{2}(\tilde{\tau}_{2}(\tau_{1}),\tilde{\tau}_{2}(\tau_{2}))\, .
\end{equation}
The double Schwarzian model can be obtain as a conformal limit of a double SYK, with fermions decoupled to one another, namely 
\begin{align}
\begin{split}
\label{HH}
\mathcal{H}\ =\ &\!\!\sum_{jklm}\!\!J^{(1)}_{jklm}\psi^{(1)}_{j}\psi^{(1)}_{k}\psi^{(1)}_{l}\psi^{(1)}_{m}+
\\
&\!\!\sum_{j'k'l'm'}\!\!\!\!J^{(2)}_{j'k'l'm'}\psi^{(2)}_{j'}\psi^{(2)}_{k'}\psi^{(2)}_{l'}\psi^{(2)}_{m'}\, .
\end{split}
\end{align}
The low-energy-limit constraint on the maps can be shown to be satisfied, since the conformal map invariance can be still preserved under time coordinates shift.

We finally remark that the $J_{1,2}$ matrix couplings are related to the double Schwarzian 
external coupling fields $f_{1,2}$, in turn related to the two dilaton fields. 

%

\section{Quantization in 1+1\emph{D}}
\noindent
The quantization of the system is ensured at the level of the SYK theory, through the maps we have recovered. 
%
Contrary to our early expectations, quantization of a multi-gravity model can be achieved just quantizing independently each single sector. This is mildly reminding what happens within the case of coupled harmonic oscillators, which can be diagonalized through a canonical transformation, and then quantized as a free harmonic oscillators. This strategy easily enables the quantization of any number of parallel models. It thus renders our construction a universal tool, simply adaptable to a system having any number of degrees of freedom. \\

More in general, we can then quantize each single sector through several equivalent methods: i) the no-boundary Hartle-Hawking wave-functions method \cite{Maldacena:2019cbz}; ii) the low-energy partition function of the double SYK model \cite{N3}; iii) the Wilson gravitational lines quantization \cite{Iliesiu:2019xuh}. In this section we provide an alternative path towards quantization in 1+1\emph{D}, as the latter strategy can be intriguingly viewed as a colored mixed Wilson lines quantization, of color mixed spin-networks. It is also worth mentioning that if the dilatons are considered as frozen, then the theory is completely topological for any possible potential. In this case, the quantization can be achieved by means of topological quantum gravity methods \cite{Dijkgraaf:2018vnm}. Nonetheless, with a conformal transformation we may take into account also dilaton degrees of freedom. In other words, the dynamical theory is located very close to the topological related version. \\

\begin{figure}[h]
\begin{center}
\begin{tikzpicture}[scale=1.8]
\draw[s]{(-1.5,0.1) -- (-1.1,0)};
\draw[s]{(-1.1,0) -- (-0.3,0.7)};
\draw[s]{(-0.2,1) -- (-0.3,0.7)};
\draw[s]{(-0.3,0.7) -- (0.3,-0.1)};
\draw[s]{(0.7,-0.1) -- (0.3,-0.1)};
\draw[s]{(0.3,-0.1) -- (-0.2,-0.6)};
\draw[s]{(-0.2,-1) -- (-0.2,-0.6)};
\draw[s]{(-0.2,-0.6) -- (-1.1,0)};
\draw[l]{(-1.4,0.9) -- (-0.3,0)};
\draw[l]{(0.4,0.7) -- (-0.3,0)};
\draw[l]{(0.4,-0.7) -- (-0.3,0)};
\draw[l]{(-1.1,-0.7) -- (-0.3,0)};
\draw[l]{(-1.4,0.9) -- (0.4,0.7)};
\draw[l]{(-1.1,-0.7) -- (-1.4,0.9)};
\draw[l]{(0.4,-0.7) -- (-1.1,-0.7)};
\draw[l]{(0.4,-0.7) -- (0.4,0.7)};
\node at (-0.75,0.52) {$s_{\varsigma}^1$};
\node at (-0.02,0.15) {$s_{\varsigma}^2$};
\node at (-0.15,-0.35) {$s_{\varsigma}^3$};
\node at (-0.65,-0.12) {$s_{\varsigma}^4$};
\end{tikzpicture}
\end{center}
\caption{Segments $s$ of the triangulation $\Delta$}
\label{bulk}
\end{figure}
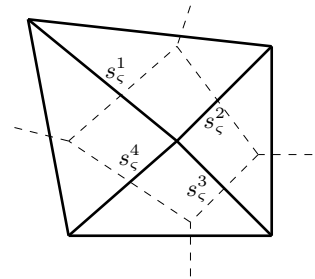
\begin{figure}[h]
\begin{center}
\begin{tikzpicture}[scale=1.8]
\draw[l]{(-1.5,0.1) -- (-1.1,0)};
\draw[l]{(-1.1,0) -- (-0.3,0.7)};
\draw[l]{(-0.2,1) -- (-0.3,0.7)};
\draw[l]{(-0.3,0.7) -- (0.3,-0.1)};
\draw[l]{(0.7,-0.1) -- (0.3,-0.1)};
\draw[l]{(0.3,-0.1) -- (-0.2,-0.6)};
\draw[l]{(-0.2,-1) -- (-0.2,-0.6)};
\draw[l]{(-0.2,-0.6) -- (-1.1,0)};
\draw[s]{(-1.4,0.9) -- (-0.3,0)};
\draw[s]{(0.4,0.7) -- (-0.3,0)};
\draw[s]{(0.4,-0.7) -- (-0.3,0)};
\draw[s]{(-1.1,-0.7) -- (-0.3,0)};
\draw[s]{(-1.4,0.9) -- (0.4,0.7)};
\draw[s]{(-1.1,-0.7) -- (-1.4,0.9)};
\draw[s]{(0.4,-0.7) -- (-1.1,-0.7)};
\draw[s]{(0.4,-0.7) -- (0.4,0.7)};
\node at (-0.3,0.2) {$\varsigma$};
\node at (-0.7,0.52) {$\varepsilon_{\varsigma}^1$};
\node at (0.15,0.3) {$\varepsilon_{\varsigma}^2$};
\node at (0.03,-0.5) {$\varepsilon_{\varsigma}^3$};
\node at (-0.65,-0.48) {$\varepsilon_{\varsigma}^4$};
\end{tikzpicture}
\end{center}
\caption{Edges $\varepsilon_{\varsigma}$ of the face $\varsigma$ of the dual complex $\Delta^*$}
\label{dual}
\end{figure}
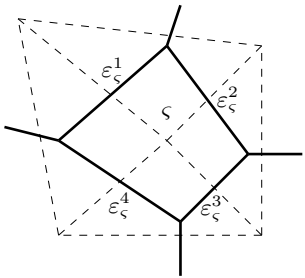

Discretization methods can be applied in the 1+1\emph{D} case, as in general to any n+1\emph{D} case. As a first step, a triangulation $\Delta$ must be picked, which introduces an arbitrary choice in the quantization procedure. Its dual $\Delta^*$ is introduced as regulator of the system. The invariance of the partition function under possible triangulations  --- namely, for a manifold $\mathcal{M}$, $\mathcal{Z}(\Delta)=\mathcal{Z}(\mathcal{M})$ --- can be either achieved accounting for regularization techniques, or exactly found in some cases. (This is for instance the case of the regularization of the Ponzano-Regge model, namely the partition-function of a BF theory, by means of the Turaev-Viro topological invariant.) As customary, frame fields are smeared along the segments $s$ of the triangulation $\Delta$ (see Fig.~\ref{bulk}), namely $B^a_s=\int_s B^a$. The field strengths of the connection $A^a$ are smeared over the faces $\varsigma$ of $\Delta^*$ (see Fig.~\ref{dual}), i.e. $F^a_\varsigma=\int_\varsigma F^a$, where the internal index $a$ is in the adjoint representation of SO$(2,1)$.
\\

As we have discussed in the previous sections, each JT sector sits very close to a topological model, sharing similarities with the theory studied by Migdal \cite{Migdal:1975zg} and Witten \cite{Witten:1991we}, which is defined by the path-integral 
\begin{equation}
\mathcal{Z}_{\rm MW}= \int \mathcal{D}A\,\mathcal{D}B \ e^{\imath \mathcal{S}_{\rm MW} [A,B] }\,,
\end{equation}
involving the 1+1\emph{D} action
\begin{equation} \label{SMW}
\mathcal{S}_{\rm MW}[A,B]=\int_\mathcal{M} {\rm Tr}[BF] +\frac{g^2}{2} \int_\mathcal{M} d\mu  {\rm Tr}[B^2]\,,
\end{equation}
with $g$ coupling constant. The action in Eq.~\eqref{SMW} does not account for the dilaton degrees of freedom. In stead, the partition function is found to depend only on the topological properties of $\mathcal{M}$, through the expression
\begin{equation}
\mathcal{Z}_{\rm MW}= \sum_j ({\rm dim}_j)^{\kappa(\mathcal{M})} e^{- g^2 \rho \, c(j)}\,,
\end{equation}
in which $j$ are unitary irreducible representations of SU$(2)$, $c(j)$ is the eigenvalue of the Casimir operator and $\rho$ the area of the dual face to the edge of the triangulation, while $\kappa(\mathcal{M})$ denotes the Euler characteristics of the manifold 1+1\emph{D} $\mathcal{M}$.\\

We are now ready to discuss the Wilson line quantization of the $BF$-like JT model for the one-sector bulk action in Eq.~\eqref{iS3}. The theory then casts 
\begin{equation}
\label{ZJT}
\mathcal{Z}_{\rm JT}\!=\!\!\int\!\mathcal{D}\phi\,\mathcal{D}A\,\mathcal{D}B\exp\!\bigg(\!\imath\!\int\! (\phi^0+\phi) \, B \wedge F[A] \bigg)\,,
\end{equation}
which corresponds, within the $BF$ formalism as in Ref.~\eqref{SMW}, to the path integral of the Einstein-Hilbert action plus a kinetic term for the scalar field, both rewritten in the Jordan frame. In this latter frame, the field-strength is conformally rescaled by the dilaton field, and thus discretization happens taking into account this term  
\begin{equation}
F^a_\varsigma(\phi) = \int_\varsigma \phi \, F^a = \phi_\varsigma\,  F^a_\varsigma \,.
\end{equation}
Adapting the procedure in Refs.~\cite{Migdal:1975zg,Witten:1991we} (see also \cite{Iliesiu:2019xuh}), one may also incorporate a generic potential $\mathcal{V}[\phi,B]$ in the action, and thus express the partition function as
\begin{equation}
\label{Z_continuous}
\mathcal{Z}_\mathcal{V}\!=\!\!\int\!\mathcal{D}\phi\,\mathcal{D}A\,\mathcal{D}B\exp\!\bigg(\!\imath\!\int\!
(\phi^0+\phi) \, B \wedge F[A]
+
\mathcal{V}[\phi,B]
\bigg).\!
\end{equation}
The role of the potential is essential for $BF$ theories (with canonical kinetic term and not accounting for any dilaton field), in order to unfreeze the degrees of freedom necessary to turn the theory from topological to metric. Nonetheless, also a deformed kinetic term may provide dynamical degrees of freedom, as we discussed in the previous sections. We inspect in details this latter case, and start our analysis setting $\mathcal{V}[\phi,B]=0$. The framework traces back to the dimensional reduction of the theory analyzed in \cite{Xu:2009bz}, further abelianized so as to account for the non-torsional limit. The discretization of the path-integral of the theory in Eq.~\ref{ZJT} then requires the introduction of group elements $g$ that are assigned to the edges $\varepsilon^n_\varsigma$ constituting the boundary of $\varsigma \in \Delta^*$ (see Fig.~\ref{dual}). This enables to recast the holonomy around $\partial \varsigma$ as
\begin{equation}
g_{\varepsilon^1_\varsigma} \circ  g_{\varepsilon^2_\varsigma} \circ \dots \circ  g_{\varepsilon^n_\varsigma} = U_\varsigma = e^{F_\varsigma}\,,
\end{equation}
where $g_{\varepsilon^i_\varsigma}=\mathcal{P}\exp \int_{\varepsilon^i_\varsigma} \! A $ is the group element associated to each edge constituting $\partial \varsigma$, and $\mathcal{P}$ denotes the path-ordering of the exponential. Summarizing, after regularization the variables of the theory are discretized as $(B^a_s, g_\varepsilon, \phi_\varsigma)$, and thus finally 
\begin{equation} \label{ZJT2}
\mathcal{Z}_{\rm JT}=\sum_\Delta \int \prod_s dB^a_s \prod_\varepsilon dg_\varepsilon \prod_\varsigma d\phi_\varsigma \, e^{-\imath  \sum_s  {\rm Tr}[ B_s (\phi_\varsigma F_\varsigma) ] } \,.
\end{equation}
In Eq.~\eqref{ZJT2} (before imposing the non-torsional limit) smeared variables are integrated in the SL$(2,\mathbb{R})$-invariant Haar measure. Invariance of the Haar measure under the M\"obius transformations for the dilaton field can be then naturally taken into account while quantizing the theory
 \begin{eqnarray}
\label{Z2}
\mathcal{Z}_f\!&=&\!\!\int\!\mathcal{D}\phi\,\mathcal{D}A\,\mathcal{D}B\exp\!\bigg(\!\imath\!\int\!
(f^0+f) \, B \wedge F[A] \bigg) \nonumber\\
&=&\!\!\int\!\mathcal{D}f\,\mathcal{D}A\,\mathcal{D}B\exp\!\bigg(\!\imath\!\int\!
(f^0+f) \, B \wedge F[A] \bigg)\,,
\end{eqnarray}
where $f:=f(\phi)$.\\

Finally, we can consider the general case of $N$-interacting JT-sectors. The kinetic term can be recovered considering M\"obius transformations
of the Haar measures that interconnect the dilaton fields $\phi_i$ of each sector. Within these assumptions, a non-trivial kinetic term can be introduced, induced by the functions $f_i:=f[\phi_i; \phi_j]$, namely 
\begin{eqnarray}
\label{Z3}
\mathcal{Z}_{\rm multi-JT}= \int \mathcal{D}\phi\,\mathcal{D}A\,\mathcal{D}B\exp\!\bigg(\!\imath\!\int\!
(f^0_i+f_i) \, B_i \wedge F[A_i] \bigg) \,. \nonumber 
\end{eqnarray}
The interacting sectors of the theory can be then decoupled using the very same invariance of the Haar measure under M\"obius transformations, and redefining the dilaton degrees of freedom, i.e.
 \begin{eqnarray}
\label{Z4} 
\mathcal{Z}_{\rm multi-JT}= \int
\mathcal{D}f\,\mathcal{D}A\,\mathcal{D}B\exp\!\bigg(\!\imath\!\int\!
(f^0_i+f_i) \, B_i \wedge F[A_i] \bigg)\,. \nonumber 
\end{eqnarray}

Finally, we consider the non-torsional limit, and incorporate the role of the potential in the discussion, by substituting the complex exponential in the path integral with 
\begin{equation}
\label{integrand}
\exp\!\bigg[\imath \!\int\!(f_i^0+f_i) \, B_i\wedge F[A_i]\bigg]\exp\! \bigg[\imath \int \mathcal{V}(B_i,B_j)\bigg]\,.
\end{equation}
We still label the edges by $\varepsilon$, but now associate to each one of them a SO$(1,1)$ group element, denoted by $g_{\varepsilon}(\eta)$, where $\eta$ stands the rapidity of the boost. Analogously, we sill denote faces by $\varsigma$, but assign to these latter irreducible representations of SO$(1,1)$, denoted as $n_{\varsigma}\in \mathds{Z}$, as quantum numbers coloring the graph. Intertwiners are trivial, and nodes do not have internal degrees of freedom in this construction. Quantizing the theory for a generic choice of the potential might be demanding. But if we pick a choice of the potential as in Eq.~\eqref{extremeflat}, quantization can be straightforwardly achieved, recovering 
\begin{equation}
\label{Z_discrete_on}
\mathcal{Z}_\mathcal{V}=\prod_{\varepsilon}\int\!dg_{\varepsilon}(\eta)
\prod_{\varsigma} \int d \phi_\varsigma
 \left( \prod \limits_{h=1}^{p} e^{\, \phi_{\varsigma}
\, n_{\varepsilon_h}
\eta_{\varepsilon_h}}
\!\right) e^{-(\sum\limits_{i} 
s^i_\varsigma )^N}.
\end{equation}
In Eq.~\eqref{Z_discrete_on}, we introduced the group measure $dg_{\varepsilon}(\eta)$; the product in the bracket is over the $p$ group elements, assigned to each edge $\varepsilon$ bounding the face $\varsigma$, and reproduces the character of the holonomy around $\varsigma$; $\phi_\varsigma$ is the discretized value of the dilaton field on each face $\varsigma$ bounded by the edges $\varepsilon_h$; the sum in the index $i$ runs over the different copies of the 1+1\emph{D} multi-gravity model; $s^i_\varsigma$ are discretized directions in the $N$-dimensional configuration space of the frame-fields for 1+1\emph{D} multi-gravity, and are assigned to each face $\varsigma$.
The quantization of the theory in Eq.~\eqref{Z2} can be achieved in a similar way as in Eq.~\eqref{Z_discrete_on}, but without accounting for the last exponential factor.

Finally, a straightforward quantization of the action on the boundary can be implemented, within the generic N-gravity framework. Focusing for the sake of simplicity the $1\!+\!1D$ scenario with only one copy of gravity, the boundary terms cast
\begin{equation}
\label{ff}
\mathcal{S}_{\partial \Sigma}=\int_{\partial \Sigma}(\phi^0+\phi^b)\, B^b 
A^b\,, \nonumber
\end{equation}
with $B^b$ and $A^b$ conjugated variables. The value of the dilaton field on the boundary can be thought as a puncture on $\partial \Sigma$, which introduces a DOF for each $1\!+\!1D$ copy of gravity. When the puncture is missing, the theory becomes topological. For a manifold provided with the Atiyah axioms, a $1D$ topological quantum field theory resembles a quantum mechanics. Indeed, to a point $p$ we may associate a Hilbert space $\mathcal{H}_p$, and then find a map that is an isomorphism among vector spaces. The action in Eq.~\eqref{ff} introduces a puncture, namely a particle, in the physical picture, hence providing near the $dS_{2}$ background solutions the equivalence with the Schwarzian quantum mechanics. 
The recovered dynamical time evolution hence individuates a unitary evolution operator.

\section{Conclusions}
\noindent
We are finally tempted to imagine that a possible approach to quantize higher dimensional gravitational theories could be inspired by a mapping into a $1\!+\!1D$ multi-gravity model, tailored in such a way to match the degrees of freedom. This might contribute to a better understanding of quantum gravity in higher dimensions. This idea certainly deserves more investigations, which are beyond the purpose of this letter. As matter of fact, a possible \textit{caveat} pertains the inevitable trivialization that a dimensional reduction would definitely encode. This issue appears to be evident once the local symmetry group structure of the gravitational theory is taken into account in higher dimensions.

\appendix

\section{BF theory and $SL(2,R)$}
\noindent 
In this section, we add a useful review on SL$(2,\mathbb{R})$ BF-theories to our previous discussion. The symmetry generators are 
\begin{equation}
\label{JA}
J_{A}=(P_{0},P_{1},P_{2})\, , 
\end{equation}
satisfying the algebraic rules
\begin{equation}
\label{PPP}
[P_{a},P_{b}]=\epsilon_{ab}\Omega, \qquad [\Omega,P_{a}]=\epsilon_{ab}P^{b}\, .
\end{equation}
The B-field and the A-field are 1-forms that can be expanded in terms 
of the symmetry generators as follows 
\begin{equation}
\label{BB}
B=t^{a}P_{a}+\phi \Omega \, ,
\end{equation}
\begin{equation}
\label{AA}
A=e^{a}P_{a}+\omega \Omega\, , 
\end{equation}
where $t_{a}$ is the torsion field, 
$\phi$ is the dilaton, 
$e^{a}$ is the zwei-bein and $\omega$ is the abelian spin connection $\omega=-\frac{1}{2}\epsilon^{ab}\omega_{ab}$.  The corresponding field-strength 
\begin{equation}
\label{FF}
F=(de^{a}+\omega \epsilon^{a}_{b}e^{b})P_{a}+\Big(d\omega +\frac{\epsilon_{ab}}{2}e^{a}\wedge e^{b}\Big)\Omega\, 
\end{equation}
\begin{equation}
  \!\!\!\! \!\!\!\! \!\!\!\! \!\!\!\! \!\!\!\! \!\!\!\! \!\!\!\! \!\!\!\! \!\!\!\! \!\!\!\! \!\!\!\! \!\!\!\! =T^{a}P_{a}+\frac{1}{2}(R-2){\rm vol}\Omega\, , \nonumber 
\end{equation}
where 
\begin{equation}
\label{dss}
d^{2}x\sqrt{-g}R=2 d\omega \, . 
\end{equation}

Thus the JT theory can be rewritten as a BF theory, namely 
\begin{equation}
\label{lal}
S_{JT}=-\frac{\phi_{0}}{8\pi}\int d\omega+\frac{1}{16\pi}\int d^{2}x \sqrt{-g}(R-2)
\end{equation}
\begin{equation}
 \!\! \!\!\!\! \!\!\!\! \!\!\!\! \!\!\!\! \!\!\!\! \!\!\!\! \!\!\!\! \!\!\!\! \!\!\!\! =\frac{\phi_{0}}{4}\chi+\frac{1}{4\pi}\int {\rm tr}(BF)\, , \nonumber 
\end{equation}
where $\chi$ is the Euler characteristic and the torsion-free constraint is imposed, turning the B-field 1-form into a scalar.

\vspace{0.3cm}
\noindent
\acknowledgments 
\noindent
The work of A.A. is supported by the Talent Scientific Research Program of College of Physics, Sichuan University, Grant No.1082204112427.
AM wish to acknowledge support by the NSFC, through the grant No. 11875113, the Shanghai Municipality, through the grant No. KBH1512299, and by Fudan University, through the grant No. JJH1512105. J.B. acknowledges his partial support by the NSFC, through the grants Nos. 11675145 and 11975203.


\end{document}